\documentclass{article}
\usepackage{ijcai25}

\usepackage{times}
\usepackage{soul}
\usepackage{url}
\usepackage[hidelinks]{hyperref}
\usepackage[utf8]{inputenc}
\usepackage[small]{caption}
\usepackage{graphicx}
\usepackage{amsmath}
\usepackage{amsthm}
\usepackage{booktabs}
\usepackage{algorithm}
\usepackage{algorithmic}
\usepackage[switch]{lineno}
\usepackage{subcaption}

\newtheorem{problem}{Problem}

\title{Peer Disambiguation in Self-Reported Surveys using Graph Attention Networks}

\author{
Ajitesh Srivastava$^1$\and
Aryan Shetty$^2$\and
Eric Rice$^3$\\
\affiliations
$^1$Ming Hsieh Department of Electrical and Computer Engineering, University of Southern California\\
$^2$Department of Computer Science, University of Southern California\\
$^3$Suzanne Dworak-Peck School of Social Work, University of Southern California\\
\emails
\{ajiteshs, asshetty, ericr\}@usc.edu
}

\begin{document}

\maketitle

\begin{abstract}
Studying peer relationships is crucial in solving complex challenges underserved communities face and designing interventions. The effectiveness of such peer-based interventions relies on accurate network data regarding individual attributes and social influences. However, these datasets are often collected through self-reported surveys, introducing ambiguities in network construction. These ambiguities make it challenging to fully utilize the network data to understand the issues and to design the best interventions. 
We propose and solve two variations of link ambiguities in such network data -- (i) which among the two candidate links exists, and (ii) if a candidate link exists.
We design a Graph Attention Network (GAT) that accounts for personal attributes and network relationships on real-world data with real and simulated ambiguities. We also demonstrate that by resolving these ambiguities, we improve network accuracy, and in turn, improve suicide risk prediction.
We also uncover patterns using GNNExplainer to provide additional insights into vital features and relationships. This research demonstrates the potential of Graph Neural Networks (GNN) to advance real-world network data analysis facilitating more effective peer interventions across various fields.

\end{abstract}

\section{Introduction}

Underserved communities such as homeless populations and veterans face higher rates of violence, substance abuse, suicide, and economic hardship~\cite{NationalAlliance2021}. A key strategy to address these issues is peer-based intervention, where individuals with firsthand experiences provide support and guidance to peers in similar situations \cite{Dennis2003}. This approach fosters trust and engagement, leading to improved intervention outcomes \cite{Solomon2004}. However, effective peer-based interventions rely on accurate network data capturing personal attributes and social influence. Incomplete or inaccurate network data can misrepresent social connections, leading to misidentification of high-risk individuals and ineffective interventions. Without reliable data, intervention strategies may target the wrong individuals, wasting resources and reducing the overall impact. A major challenge in gathering such data is the ambiguity in self-reported ``social network'' surveys.  When asked to list peers, the participant only mentions names. There may be multiple individuals of the same name; or there may be someone in the dataset with that name but the participant was referring to someone else. These ambiguities create unreliable graphs, limiting the predictive power of machine learning models and the effectiveness of interventions. Following up with the participants after the full round of data collection to resolve these ambiguities is time-consuming and difficult.

This ambiguity hinders the accuracy of methods for predictions and reduces the effectiveness of intervention strategies, leading to reliance on traditional methods. Conventional approaches to resolve ambiguities rely on subjective human coding, where multiple coders attempts to identify the correct link by considering the attributes of an individual.
While this approach may be effective, it is time-consuming, inconsistent, and difficult to scale. We aim to develop a framework that efficiently resolves these ambiguities while providing researchers with interpretable explanations for data refinement. Once the ambiguities are resolved, we can use the network data to build more reliable models (for predictions or to understand behavior dynamics) that can then be used for peer interventions.

This work is part of a project to improve the mental health of at-risk populations. These methods also apply to our other projects on substance use and violence prevention among youth experiencing homelessness. As such, it aligns with SDG 3 (Good Health and Well-Being) by enhancing data-driven intervention strategies that promote mental health support and crisis prevention for vulnerable groups. This work is a collaboration between engineering and social work alongside community partners in Los Angeles\footnote{Names omitted to preserve anonymity}. The collaboration ensures that our approach is grounded in real-world intervention strategies, making AI-driven solutions more applicable and impactful.

We propose using Graph Attention Networks (GATs) to model social interactions and resolve ambiguities by predicting which interactions are likely. We address two key problems: (i) Pair disambiguation, where multiple individuals share the same reported name, and (ii) Link existence, where a reported connection may or may not exist in the network. By addressing these challenges, we enable machine learning-based targeted interventions, ultimately improving the accuracy of network-based intervention strategies. As a demonstration, we develop methods for predicting suicide risk in an active military population and show that the performance improves after resolving ambiguities.

\subsection{Background and Related Work}

\subsubsection{Peer-based Interventions}

By leveraging social connections, peer-based interventions can empower individuals to act as agents of change. These interventions are seen in the form of sharing knowledge and encouraging behaviours such as HIV prevention, substance abuse reduction, and mental health support.

Peer-based intervention is an ideal mechanism for underserved communities as they are often wary of authority figures, institutions, and organizations, including social service agencies
and health care clinics \cite{fest2013street}. This wariness is prompted by factors including stigma and discrimination surrounding the community \cite{martins2008experiences} and feelings of distrust in this population fostered by past experiences in the formal service system \cite{dworsky2009homelessness}. Therefore, these individuals are more likely than other populations to not seek help from traditional service providers, choosing instead to rely on informal sources such as friends and media \cite{malow2007hiv}.

\subsubsection{Challenges in Data Collection and Ambiguity}
 Effective data collection is vital for the success of network-based interventions; however, it is frequently hindered by the inherent ambiguities and inaccuracies of self-reported data. Respondents may submit incomplete or biased responses, resulting in unreliable network information. Studies by Marsden \cite{Marsden1987} and Freeman \cite{Freeman1987} outline the limitations of self-reported surveys and the ensuing difficulties in social network analysis. These challenges emphasize the necessity for advanced techniques to minimize data ambiguity and enhance the reliability of network evaluations. 
 Beyond name ambiguity, self-reported data introduces inconsistencies due to recall bias, reluctance to disclose sensitive connections and misunderstandings in reporting. Participants may not fully disclose their relationships, leading to incomplete or distorted network structures. These challenges complicate traditional validation methods, making an automated approach to resolving ambiguities in network data essential.
 
\subsubsection{Graph Neural Networks in Social Network Analysis}
Graph Neural Networks (GNNs) effectively analyze complex networks by leveraging relational structures to uncover patterns missed by traditional methods. Kipf \cite{Kipf2017} introduced Graph Convolutional Networks (GCNs) for link prediction and node classification, while Velickovic \cite{Velickovic2018} proposed Graph Attention Networks (GATs) to enhance social network modeling with attention mechanisms. While GNNs have been widely applied in social network analysis, their use in resolving self-reported survey ambiguities remains unexplored. Our work bridges this gap by leveraging GATs to refine network representations, improving the accuracy of interventions.

\subsection{Contributions}

\begin{enumerate}
    \item We develop a Graph Attention (GAT)--based framework to assist social scientists with link disambiguation in the construction of network data.
    \item We evaluate our approach on a real dataset of active-duty military personnel where real ambiguity was observed and resolved by human coders. We also evaluate our approach by simulating ambiguity on this dataset.
    \item We demonstrate through experiments on both real and simulated ambiguity that our GAT-based approach outperforms the baselines in almost all metrics for disambiguation.
    \item We evaluate the impact of disambiguation by testing it along with a task of predicting suicide risk.
    We demonstrate that suicide risk prediction accuracy improves after addressing the ambiguities in the network.
    \item We provide an explanation of the disambiguation model using GNNExplainer in terms of relevant features and neighborhoods. This is crucial in Social Sciences for assisting human coders in understanding the reasoning behind relationships in the network.
\end{enumerate}

\section{Data and Problem Definition}

\subsection{Dataset Description}

We use a dataset~\cite{barr2025supports} collected from a group of 250 active-duty military personnel in Los Angeles, capturing various demographics and service-related features. This dataset contains extensive and diverse information on multiple aspects of veterans' lives, such as their demographic profiles, educational backgrounds, and military experiences. The list of attributes is given in Table~\ref{tab:dataset_attributes}. 

\begin{table}[!t]
\centering
\caption{Dataset Attributes Description}
\label{tab:dataset_attributes}
\begin{tabular}{|c|p{2.2in}|} \hline

\textbf{Attribute} & \textbf{Description} \\ \hline

PID & Unique identifier for each veteran \\ \hline
Age & The participant's age \\ \hline
Race & The participant's race\\ \hline
ed & The participant's educational 
 attainment\\ \hline
marital\_status & The participant's marital status  \\ \hline
Romantic & Binary indicator of whether the participant 
is currently in a romantic relationship \\ \hline
Gender & The participant's gender \\ \hline
Sexuality & The participant's sexuality\\ \hline
Siblings & The count of the participant's siblings \\ \hline
Household & The total number of individuals in the 
\\ & participant's household \\ \hline
dad ed& The educational attainment of the 
\\ & participant's father \\ \hline
mom ed& The educational attainment of the 
\\ & participant's mother \\ \hline
military\_join & The year the participant enlisted in 
\\ & the military \\ \hline
Rank & The participant's military rank \\ \hline
MOS & The Military Occupational Specialty 
\\ &(MOS) code assigned to the participant \\ \hline
deploy ever& Binary indicator of whether the participant 
\\ &has ever been deployed \\ \hline
Deployments & The total number of deployments 
\\ & experienced  by the participant \\\hline
deploy where& The locations of the participant's deployments \\ \hline

\end{tabular}
\end{table}

\begin{figure*}[!h]
    \centering
    \includegraphics[width=0.75\linewidth]{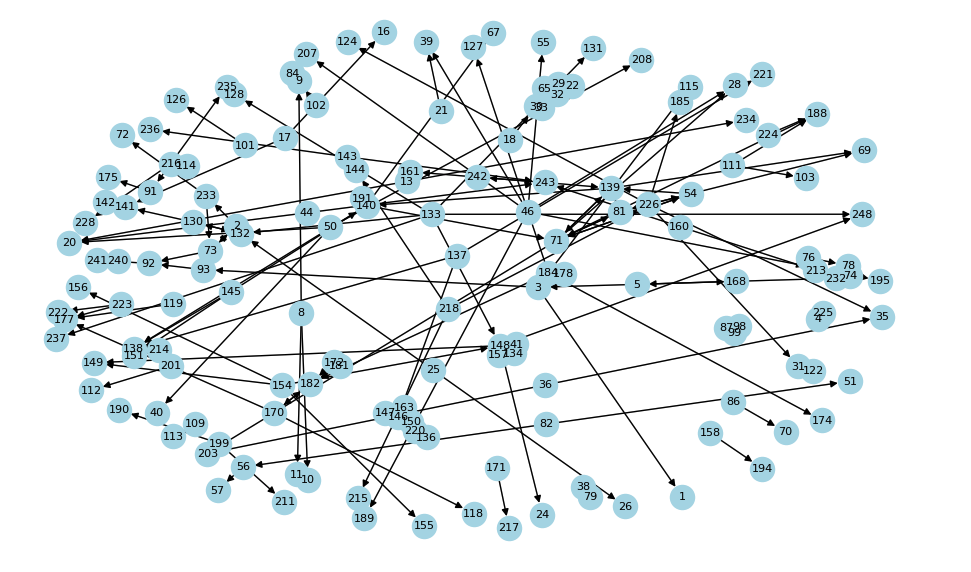}
    \caption{Graph network from the dataset}
    \label{fig:network}
\end{figure*}

Along with the listed attributes, this dataset contains network information obtained from a social network interview.  It asked participants to list up to 10 people with whom they had contact in the past 30 days, including those they take advice from and those who would listen if they needed to talk. These are used to generate the network structure. The network obtained from this dataset is shown in Figure~\ref{fig:network}. Each individual is a node in the network. Based on the listed contacts, two human coders identify the ``alters'' (nodes that the current individual should be connected to ) in the dataset. There may be multiple people with the same name. Also, the mentioned name may be of an individual outside the dataset but is shared by someone in the dataset. The human coders attempt to resolve the ambiguities and assign a confidence score. For some links there is high confidence, while for some, the confidence is low. We treat the high confidence links as our ground truth. To summarize, the network has 242 nodes, 275 total edges of which 184 are confident edges and the rest are low confidence as annotated by human coders. Each node has a large of attributes including the listed attributes, and responses specific to mental health. Only the listed attributes were used for disambiguation as they are objective attributes that were considered by the human coders to construct the edge list.

\subsubsection{Disambiguation Problems}

Motivated by the network construction process challenges, we define two problems.

\begin{problem}[\textbf{Pair Disambiguation}] Given a node $u$ and its features, and those of two other nodes $v_1$ and $v_2$, decide which one of the two links $(u, v_1)$ and $(u, v_2)$ exists.
    
\end{problem}

\begin{problem}[\textbf{Link Existence}]
    Given two nodes $u$ and $v$ and their features, decide if the link $(u, v)$ exists.
\end{problem}

By resolving these uncertainties, we enhance network accuracy thus ensuring that interventions reach the right individuals based on actual social relationships rather than potentially erroneous self-reported data. 

\section{Method}

\begin{figure*}[!ht]
    \centering
    \includegraphics[width=1\linewidth]{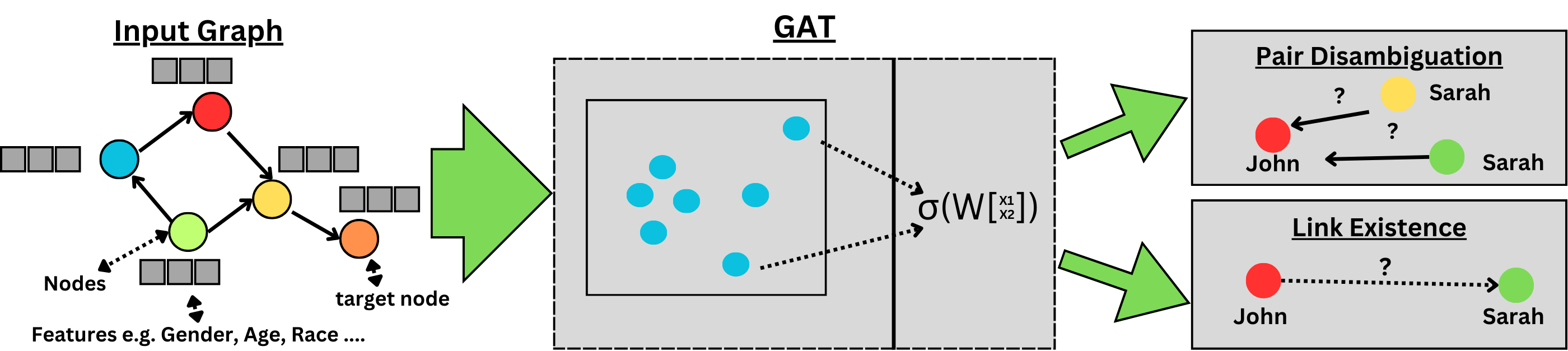}
    \caption{Approach overview: The input graph consisting of nodes representing individual features is fed into the GAT model to generateembeddings.  These embeddings are used to resolve the ambiguity between multiple candidates (pair disambiguation) and to determine the likelihood of a link between individuals (link existence).}
    \label{fig:overview}
\end{figure*}

We leverage a Graph Attention Network (GAT), which takes preprocessed features as inputs and generates an embedding. The preprocessing involves transforming the categorical features into a set of binary features, normalizing numerical features, and removing features with a majority of values missing. The output embedding aims to have a small distance between nodes that are connected and a high distance between those that are not.
The GAT consisting of several attention layers, is trained on links with no ambiguity. The embeddings generated by this model form the basis of our solution to the above-mentioned problems. For Pair Disambiguation, we use these embeddings to help determine which of the two candidate nodes is the alter for the given nopde. In the Link Existence problem, the embeddings assist in predicting whether a link exists between specific pairs of nodes based on how close the nodes are. An overview of the approach is given in Figure~\ref{fig:overview}. To interpret the model's predictions, we employ the GNNExplainer, which provides insights into the key features and edges that affect the prediction results. By analyzing subgraphs around chosen node pairs and emphasizing important edges and features we can understand the model's decision-making process. 

\subsection{Embedding Model: Graph Attention Network (GAT)}

The Graph Attention Network (GAT) model is used to create embeddings for nodes, which forms the basis of our solution to pair disambiguation and link existence. This model leverages attention mechanisms to dynamically combine features from the neighbouring nodes resulting in strong representations that encompass the structural and attribute information within the graph \cite{Velickovic2018}.

For every node, the model determines attention coefficients that reflect the significance of adjacent nodes. These attention coefficients are calculated using:
\[
e_{ij} = \text{LeakyReLU}\left(a^T \left[ W \mathbf{h}_i \| W \mathbf{h}_j \right]\right)
\]
where \( \mathbf{h}_i \) and \( \mathbf{h}_j \) are the feature vectors of nodes \(i\) and \(j\), \(W\) is the weight matrix, \(a\) represents the attention vector and \(\|\) denotes concatenation. These coefficients are then normalized using the softmax function:

\[
\alpha_{ij} = \frac{\exp(e_{ij})}{\sum_{k \in \mathcal{N}(i)} \exp(e_{ik})}
\]
Here, \( \alpha_{ij} \) represents the normalized attention coefficient for the edge between nodes \(i\) and \(j\), and \(\mathcal{N}(i)\) denotes the set of neighbours of node \(i\). 

\paragraph{Feature Aggregation:}

We aggregate features using the attention coefficients. For each node, the updated feature vector \( \mathbf{h}_i' \) is calculated as:
\[
\mathbf{h}_i' = \sigma \left( \sum_{j \in \mathcal{N}(i)} \alpha_{ij} W \mathbf{h}_j \right)
\]
where \( \sigma \) is a non-linear activation function like ReLU. This aggregation process allows the model to incorporate information from the node's neighbours weighted by their importance.

\paragraph{Stacking GAT Layers}
To capture higher-order neighbourhood information we stack multiple GAT layers. Each layer applies the attention mechanism and feature aggregation, with the output of one layer serving as the input for the next. This hierarchical approach enables the model to learn more complex patterns and interactions within the graph.

\paragraph{Model Architecture and Training}

Our GAT model consists of several convolutional layers. The architecture includes: (i) an initial GATConv layer with 16 hidden channels and 8 attention heads;
(ii) one additional GATConv layer, where the input dimension (16 × 8 = 128) is processed through 8 attention heads; and 
(iii) a final GATConv layer producing 7 output channels using a single attention head

The exact numbers of GAT layers and channels are determined through a validation set to pick the best-performing architecture. The model is trained using backpropagation to minimize the binary cross-entropy loss between predicted and actual link existence. The binary cross-entropy loss function is given by:

\[
\mathcal{L} = -\frac{1}{N} \sum_{i=1}^N \left[ y_i \log(\hat{y}_i) + (1 - y_i) \log(1 - \hat{y}_i) \right]
\]
where \(y_i\) denotes the true label (1 for existing links, 0 for non-existing links), and \(\hat{y}_i\) represents the predicted probability of a link between nodes \(i\) and \(j\). 
Pair disambiguation can be addressed by checking which of the two links is more likely to exist. 

\subsection{Evaluation Setup}
\subsubsection{Real Ambiguity} 
Recall that our dataset contains confidence scores associated with links that distinguish those with no ambiguity and those with ambiguities that were resolved. We train our model based on the former group (no ambiguity) of links, and we test the model on the latter group. There were 4 pairs of ambiguous links that were resolved by human coders. These fall under the Pair Disambiguation problem. Given the pairs where ambiguity existed, we compare the distance in the embedding of the source node with each of the two nodes. The node with a smaller distance is chosen. Further, there were 83 links potentially referring to individuals outside the network. These fall under the Link Existence problem. If the distance of the given node from the source node in the embedding is lower than a threshold, then we declare that the link exists. The threshold is selected based on a validation set.

\subsubsection{Simulated Ambiguity} 
Given the small number of real examples for both problems, we also simulated these problems on the high-confidence links. We split the high-confidence links into a train, a validation, and a test split of 60\%, 20\%, and 20\%, respectively. On the test set, we simulate both types of ambiguities. For Pair Disambiguation, we randomly sample (with replacement) three nodes $u, v_1,$ and $v_2$ such that the link $(u, v_1)$ exists in the test set and $(u, v_2)$ does not exist in the entire graph. The task is to be able to correctly identify $(u, V_1)$ exists. For Link Existence, we randomly select nodes $u$ and $v$ from the test set, and the task is to identify if the link $(u, v)$ exists. The links selection is made so that the positive (actual link) and negative samples (no link) are balanced.

\section{Experiments}

We removed features with a majority of null values. We encoded categorical features like military rank and romantic status to sets of binary features and normalized the continuous features. 
To evaluate the models' performance we use accuracy, precision, recall, F1 score, and accuracy. Our code is publicly available\footnote{\url{https://github.com/scc-usc/peer-disambiguation}}.

\subsubsection{Baselines}
\noindent\textbf{Decision Trees:}
We used a decision tree classifier as a baseline. We prepared the data by concatenating the features of both the source and destination nodes of positive (link present) and negative (link absent) edge samples. We then trained the decision tree model on these samples. Due to the interpretable nature of the decision tree, we can also draw insights into the importance of various features.

\noindent\textbf{Multilayer Perceptron (MLP)}
We used a Multilayer Perception (MLP) as a baseline. We prepared the data by concatenating the features of both the source and destination nodes of positive (link present) and negative (link absent) edge samples. We then trained the MLP on these samples. MLP provides a non-graph-based comparison that assesses the link prediction solely based on node attributes. 

\subsection{Disambiguation Results}

\begin{table}[H]
    \centering
    \begin{tabular}{|c|cccc|} \hline
         Model & Precision & Recall & F1 & Accuracy \\  
         \hline
         Decision Tree & 1.0 & 0.39 & 0.56 & 0.39\\ 
         MLP &  0.90 & 0.64 & 0.74& 0.78\\
         GAT &  \textbf{1.0} & \textbf{0.93}& \textbf{0.96}& \textbf{0.93}\\\hline
    \end{tabular}
    \caption{Results for Simulated Pair Disambiguation}
    \label{tab:link_disambiguation}
\end{table}

\begin{table}[H]
    \centering
    \begin{tabular}{|c|cccc|} \hline 
         Model & Precision & Recall & F1 & Accuracy \\ \hline
         Decision Tree & 1.0 & 0.5 & 0.6667 & 0.5 \\  
         MLP &  1.0 & 0.5 & 0.6667 & 0.5 \\
         GAT &  \textbf{1}& \textbf{1}& \textbf{1}& \textbf{1}\\\hline
    \end{tabular}
    \caption{Results for Real Pair Disambiguation}
    \label{tab:real_node_disambiguation}
\end{table}

\begin{table}[H]
    \centering
    \begin{tabular}{|c|cccc|} \hline 
         Model & Precision & Recall & F1 & Accuracy  \\ \hline
         Decision Tree &  0.65& 0.89& 0.75& 0.71\\  
         MLP &  0.68 & \textbf{0.90} & 0.77& 0.74 \\
         GAT &  \textbf{0.94}& 0.66 & \textbf{0.77}& \textbf{0.80}\\\hline
    \end{tabular}
    \caption{Results for Simulated Link Existence}
    \label{tab:link_existence}
\end{table}

\begin{table}[H]
    \centering
    \begin{tabular}{|c|cccc|} \hline 
         Model & Precision & Recall & F1 & Accuracy \\ \hline
         Decision Tree & \textbf{1.0} & 0.63 & 0.77 & 0.63\\  
         MLP &  1.0 & 0.42 & 0.59 & 0.42\\
         GAT &  0.77 & \textbf{1.0} & \textbf{0.87}& \textbf{0.77}\\\hline
    \end{tabular}
    \caption{Results for Real Link Existence}
    \label{tab:real_link_existence}
\end{table}

Tables~\ref{tab:link_disambiguation},~\ref{tab:real_node_disambiguation},~\ref{tab:link_existence}, and ~\ref{tab:real_link_existence} show the results. 

GAT significantly outperforms the baseline models for both problems across most evaluation metrics for the simulated and real applications. In the pair disambiguation simulation, the GAT model outperforms the baselines in all metrics with an accuracy of \textbf{0.93} for the simulated link disambiguation and a perfect score of 1 for the limited real examples of pair disambiguation. These results exceeded those of the baselines as MLP came at \textbf{0.64} and \textbf{0.67} respectively.

For the simulated and real link existence, it achieves the highest accuracy of \textbf{0.81} on the simulated link existence and \textbf{0.77} on the real examples of link existence checking compared to the other closest baseline coming in at \textbf{0.74} accuracy and \textbf{0.63} respectively.

The simulations and real examples verify that the embeddings generated by the GAT model accurately represent node relationships within the graph, providing a much more reliable representation of the social network. 

\subsection{Application: Suicide Risk Prediction}

To test the impact of link disambiguation, we consider the task of predicting the suicide risk of each individual in the network.  
First, we apply our GAT model to resolve ambiguous connections by predicting which potential alter (connection) is correct for each uncertain edge. If a PID has multiple possible alters, the model determines the correct match using link existence predictions or embedding distances. A total of 79 new edges were added, increasing the edge count for confident edges from \textbf{184} to \textbf{263}. Recall that the original edge list provided by human coders included 184 confident links and 91 uncertain links (total 275).

To train the prediction model, we include 271 additional node attributes available in the dataset. These additional attributes capture various aspects of mental health, including childhood experiences, depression, and PTSD. While these attributes could have been included during disambiguation, we excluded them as they are specific to the prediction task. Additionally, the dataset includes four direct indicators of suicidal intention that are used to compute a suicide score which is treated as the prediction target. 

We compute the suicide score based on direct indicators of suicidal ideation, summing responses and applying sigmoid normalization. Following the SBQ-R manual, we use a validated cutoff score of 7 to classify individuals at elevated suicide risk \cite{gutierrez2019psychometric}. This score serves as a proxy for identifying individuals who may require intervention, making it crucial to ensure that it is derived from an accurate representation of social ties and risk factors. Since the dataset had a limited number of at-risk individuals (high suicide score), we oversample them. This ensured a balanced representation, allowing the models to learn a fair distribution across the range of the score. 

We train a GAT regression model on the updated dataset and disambiguated edge list to predict suicide scores for each individual, incorporating both node attributes and network structure for a more precise risk assessment. This provides a comparison of performances between the original and disambiguated edge lists.
MAE quantifies the prediction accuracy of the regression models, while AUC measures the model’s ability to distinguish at-risk individuals by comparing predicted scores against the cutoff. We apply the same evaluation to a Decision Tree and an MLP trained on the updated dataset and edge list to assess their performance relative to the GAT model.

\begin{table}[h]
    \centering
    \begin{tabular}{|c|cc|}
        \hline
        Model & MAE & AUC \\ \hline
        Decision Tree on Original Edge List & 1.89 & 0.74 \\ 
        Decision Tree on Disambiguated Edge List & \textbf{1.36} & \textbf{0.84} \\ \hline
        MLP on Original Edge List & 1.67 & 0.58 \\ 
        MLP on Disambiguated Edge List & \textbf{1.13} & \textbf{0.75} \\ \hline
        GAT on Original Edge List & 0.19 & 0.97 \\ 
        GAT on Disambiguated Edge List & \textbf{0.14} & \textbf{0.99} \\ \hline
    \end{tabular}
    \caption{Comparison of Model Performance on Suicide Score Prediction with Original and Disambiguated Edge Lists}
    \label{tab:suicide_score_comparison}
\end{table}

Table \ref{tab:suicide_score_comparison} shows the results of the predictions on original edge list vs disambiguate edge list. 
Resolving network ambiguities using our proposed method improved suicide risk prediction, reducing MAE from \textbf{0.19} to \textbf{0.14} and increasing AUC from \textbf{0.97} to \textbf{0.99}. In fact, even non-GNN models benefited from the disambiguated edge list, thus demonstrating that resolving ambiguities in network connections can significantly improve the predictive power of machine learning models. This improved structure enables more precise identification of at-risk individuals, minimizing false positives and negatives, and strengthening peer-based interventions.

Beyond suicide risk, improving network accuracy has broader implications for network-driven interventions for various applications, such as substance use prevention and crisis response. Refining relationships within the social network enhances data reliability, which is crucial for designing targeted interventions in high-risk communities. Structural improvements in the network, such as increased connectivity and reduced spurious links, suggest that disambiguation methods can improve predictive models across multiple domains where relational data plays a key role.

\section{Discussion}

We wish to generate explanations that can be provided to the user to help develop the rationale behind the connections based on node attributes and interactions. 

\subsection{Explanation}

\subsubsection{Understanding GAT behavior}
\begin{figure}[!ht]
    \centering
    \begin{subfigure}[]{0.75\linewidth}
        \includegraphics[width=\linewidth]{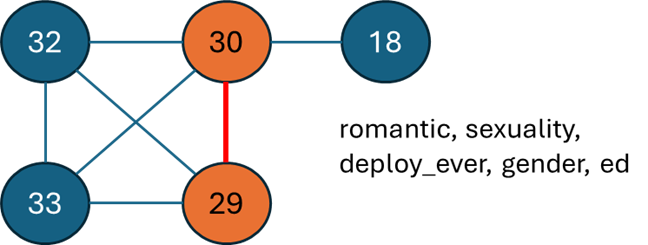}    
        \caption{Informative subgraph}
        \label{fig:good_sub}
    \end{subfigure}
    \begin{subfigure}[]{0.75\linewidth}
        \includegraphics[width=\linewidth]{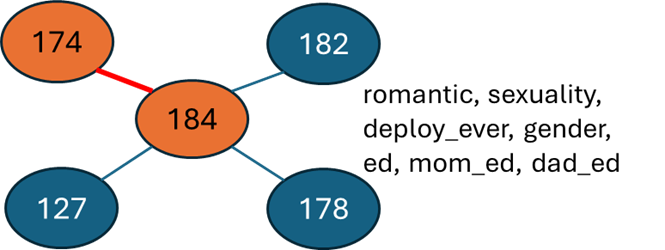}  
        \caption{Uninformative subgraph}
        \label{fig:bad_sub}
    \end{subfigure}
    \caption{Subgraph and features identified by GNNExplainer as decisive in identifying the existence of link given in red. (a) The nodes share common neighbors and a few important features. (b) The nodes do not share any common neighbors, but many important features}
    \label{fig:enter-label}
\end{figure}

We examined the outcomes of link existence between various node pairs using our GAT model, with insights provided by GNNExplainer~\cite{ying2019gnnexplainer}. 
It attempts to find a subset of features and subgraphs that are critical in determining the model's behavior (predicting the likelihood of links). We load the pre-trained GAT model and configure the GNNExplainer for binary classification at the edge level. GNNExplainer finds a subgraph and subset of features that contain most information in making the decision through our model. 
The results demonstrated consistent trends across different node pairs highlighting the features that impact the model's ability to determine if a link exists. 
\noindent\textbf{Node Features: } For all node pairs analyzed, source and destination nodes frequently had shared attributes in gender, romantic status, and marital status. 
    Characteristics related to race, education, and parental education, were less frequent but often emphasized. 
    Rank, deployment history, and military occupational speciality (MOS) were considered important by GNNExplainer but frequently differed between the pair of nodes. 
\noindent \textbf{Subgraph: } In some cases, as shown in Figure~\ref{fig:good_sub}, having a shared neighborhood strongly influenced the link existence. However, there were cases where no neighborhood was shared in the subgraph considered important (Figure~\ref{fig:bad_sub}). In such cases, more shared features appeared to be important. Therefore, node features or subgraphs alone were not sufficient for the decision.  

\subsubsection{Decision tree splits}
\begin{figure}
    \centering
    \includegraphics[width=1\linewidth, trim={0, 2.8cm, 0, 0}, clip]{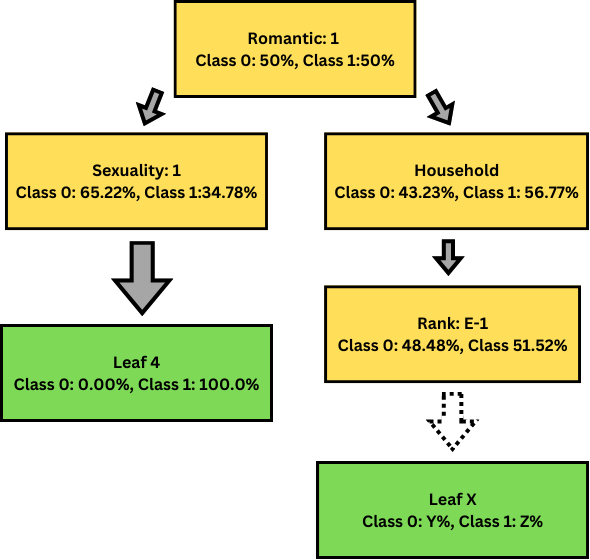}
    \caption{Decision Tree with class distribution}
    \label{fig:tree}
\end{figure}

We analyzed the early features used in the decision tree. A part of the tree is shown in Figure~\ref{fig:tree}. These features turned out to be romantic status, sexuality, household, military rank, demographics, and education. The majority of these overlap with the features identified by the GNNExplainer on our GAT model. The similarities between decision trees and GAT suggest that these features are essential for confirming links between nodes. However, there are differences in the order of importance across the two methods. While both methods recognize similar key features, they rank them differently.

\subsubsection{Intepretability for Decision-Making}

Network-based AI models must be transparent, especially in sensitive areas like suicide prevention. Our approach ensures that the researchers and agencies can understand how ambiguous connections are resolved rather than relying on the model without scrutiny. By clarifying the decision-making process, we can further improve the effectiveness and trustworthiness of network-based interventions.

\subsubsection{Homophily in Social Networks}
Research on homophily in social networks demonstrates that shared attributes such as gender and romantic status significantly increase the likelihood of a connection existing \cite{McPherson2001}. This effect is also seen in professional networks where attributes like race, education and parental education impact connections as supported by \cite{Marsden1987}. Demographic and professional information including race and education play a crucial role in shaping network connections as seen in \cite{Lizardo2006} where the findings concluded the influence of cultural tastes and social identity in network formation. 


\section{Conclusions}

We addressed difficulties related to network data construction in research on underserved communities. Specifically, we identified and solved disambiguation in identifying alters (the person a given individual is referring to) in a social network survey. We developed a GAT-based approach that first generates an embedding of the network based on node features and unambiguous links. We used these embeddings to identify the correct alter.
We demonstrated through experiments that our approach outperforms a MLP method and a decision tree. We also demonstrated that improving the quality of the network data through disambiguation improved the performance of machine learning models on the task of predicting suicide risk.
Using GNNExplainer, we identified useful features and subgraphs decisive in the embedding. We envision that our method along with the explanation will assist social science researchers in accelerating their data construction process. This work strengthens the foundation for AI-driven social interventions, improving resource allocation and outreach in vulnerable communities
Having unambiguous data will enable more effective peer interventions in underserved communities as having cleaner data will enable more machine learning-based methods in these interventions. 



\clearpage

\bibliographystyle{named}

\end{document}